# Insights into the LiMn$_2$O$_4$ Cathode Stability in Aqueous Electrolyte

Juan Carlos Gonzalez-Rosillo[a*], Maxim Guc[a], Maciej Oskar Liedke[c], Maik Butterling[c], Ahmed G. Attallah[c], Eric Hirschmann[c], Andreas Wagner[c], Victor Izquierdo-Roca[a], Federico Baiutti[a], Alex Morata[a*], Albert Tarancón[a,b*]

[a] Catalonia Institute for Energy Research (IREC), Jardins de les Dones de Negre 1, Planta 2, 08930, Sant Adrià del Besòs, Barcelona, Spain

[b] Catalan Institution for Research and Advanced Studies (ICREA), Passeig Lluís Companys 23, 08010, Barcelona, Spain

[c] Helmholtz-Zentrum Dresden - Rossendorf, Institute of Radiation Physics, Bautzner Landstraße 400, 01328 Dresden, Germany

*Corresponding authors

LiMn$_2$O$_4$ (LMO), cathodes present large stability when cycled in aqueous electrolytes, contrasting its behavior in conventional organic electrolytes in Lithium-ion batteries (LIBs). To elucidate the mechanisms underlying this distinctive behavior, we employ unconventional characterization techniques, including Variable Energy Positron Annihilation Lifetime Spectroscopy (VEPALS), Tip-Enhanced Raman Spectroscopy (TERS) and macro-Raman Spectroscopy (with hundreds of µm-size laser spot). These still rather unexplored techniques in the battery field provide complementary information across different length scales, revealing previously hidden features.

VEPALS offers atomic-scale insights, uncovering cationic defects and sub-nanometer pores that tend to collapse with cycling. TERS, operating at the nanometric range at the surface, captured the presence of Mn$_3$O$_4$ and its dissolution with cycling, elucidating dynamic changes during operation. Additionally, TERS highlights SO$_4^{2-}$ accumulation at grain boundaries. Macro-Raman Spectroscopy focuses on the micrometer scale, depicting small changes in the cathode's long-range order, suggesting a slow but progressive loss of crystalline quality under operation.

Integrating these techniques provides a comprehensive assessment of LMO cathode stability in aqueous electrolytes, offering multifaceted insights into phase and defect evolution that can help to rationalize the origin of such stability when compared to conventional organic electrolytes. Our findings advance the understanding of LMO behavior in aqueous environments and provide guidelines for its development for next-generation LIBs.

## Introduction

The development of advanced cathode materials for lithium-ion batteries (LIBs) has been at the forefront of research aiming at improving the performance, cost and safety of energy storage systems. Since its first reports by J.C. Hunter and Goodenough's group back in the 80s,[1,2], there has been a continuous and growing interest from both academic and industrial perspectives in $LiMn_2O_4$ (LMO), which has led to a deep understanding of its structure, properties, advantages and challenges.[3] LMO belongs to the family of cubic spinel oxides (F$d$3$m$ space group), in which Li is tetrahedrally coordinated and Mn is octahedrally coordinated, which allows for 3D $Li^+$ diffusion channels that could potentially lead to fast-charging cathodes, as shown, for instance, in other spinel-type materials.[4]

Despite its relatively high theoretical capacity (148 and 296 mAh·g$^{-1}$ for $LiMn_2O_4$ in the regular and in the extended cycling range, respectively), there are plenty of reports showing concerns regarding the long-term stability when cycled in conventional organic electrolytes. In particular, Mn dissolution has been claimed to be the main drawback of this material, partially due to Jahn-Teller distortions and Mn disproportionation.[5] However, other effects play a role, since Mn dissolution also occurs in the fully charged state,[6,7] unexpected if disproportionation would be the only responsible of the $Mn^{2+}$ dissolution. Recent works have shown with great detail the degradation mechanism in $LiMn_2O_4$ in LIBs through advanced characterization techniques, as excellently summarized by Prof. Ju Li's team in their review.[3] Essentially, these recent studies have highlighted significant surface-related phenomena. These observations include Mn reduction during charging and Mn oxidation during discharging at the surface, oxygen loss linked to Mn reduction, and surface reconstruction upon charging.[8–11] This ultimately results in Mn dissolution and surface phase transformations, contributing to capacity decay and increased impedance.[12] In addition, advanced characterization using in situ techniques has revealed the formation of $Mn_3O_4$ at high charge voltages of 4.2V and $Li_2Mn_2O_4$ at discharge voltages of 3.3V. These electrochemically induced phase transformations are partially irreversible and can lead to particle cracking during cycling, further exacerbating the reactive surface area.[13]

Interestingly, despite the intricate challenges posed by surface-related phenomena and structural instability in $LiMn_2O_4$ when employed with conventional organic electrolytes, a remarkable shift occurs when this cathode material is cycled in aqueous electrolytes. In this alternative environment, $LiMn_2O_4$ exhibits a surprising level of stability, seemingly untouched by the issues that affect its performance in organic electrolytes, besides the well-known Mn dissolution.[14–19] This stark contrast in behavior underscores the need for a thorough investigation into the underlying mechanisms responsible for this stability divergence, setting the stage for our exploration of LMO's performance in aqueous environments using unconventional techniques, trying to ask a simple question: why is LMO more stable in aqueous electrolytes? Can we learn anything from the study of high-stability operation in aqueous electrolyte that can be extrapolated to organic electrolytes?

In this work, we examine the structural and phase evolution of LMO films in aqueous electrolyte as a function of cycling by three novel characterization techniques that allow us to examine surface and bulk properties at very different length scales: Tip-Enhanced Raman Spectroscopy (TERS), macro-Raman Spectroscopy and Variable Energy Positron Annihilation Lifetime Spectroscopy (VEPALS).

TERS is a relatively new surface-sensitivity technique that overcomes the spatial resolution limitations of conventional Raman (~500 – 1000 nm) reaching even single molecule detection.[20–23] While it is widely used to study organic materials, its application to complex oxides is not yet widespread. The use of TERS in batteries is still in its infancy.[24,25] Essentially, TERS combines the spectroscopic power of Raman Spectroscopy with the spatial resolution of a Scanning Probe Microscope. This is achieved by carefully focusing the laser at the apex of a tip coated with a thin metal able to generate surface plasmon resonances and lightning-rod effect to enhance the Raman signal.[26]

With a completely different and complementary approach, macro-Raman Spectroscopy uses a very large laser spot size (in the range of 50 – 100 µm in diameter) and ultra-low energy densities (tens of W·cm$^{-2}$ vs. tens of kW·cm$^{-2}$ in conventional Raman) to ensure no influence on the sample while being sensitive to phases located in a true representative area, with the averaged signal from $10^3$-$10^4$ grains under the measurement spot.

In VEPALS, the films are irradiated with positrons that recombine (annihilate) with electrons in the films at defined depths, creating gamma rays, which are subsequently detected.[27] The measurement of positrons lifetime provides a direct information on the type and size of negatively charged and neutral defects through their characteristic time before encountering an electron, whereas pores are characterized based on the lifetime of residing and bouncing in them a bound state of positron and electron, the so-called Positronium (Ps) [28,29]. The concentration of positrons and Ps is evaluated through the relative intensities of the different lifetime components. In the context of battery research, VEPALS has found notable applications, particularly in studies involving layered compounds like $LiCoO_2$. In such cases, positron lifetime has been proposed as a reliable indicator of the lithiation state of grain boundaries.[30] Furthermore, in investigations of compounds like $LiNi_{1/3}Mn_{1/3}Co_{1/3}O_2$, operando experiments have revealed a gradual increase in divacancy and vacancy agglomerate formation during the charging process, culminating in the transformation of these agglomerates into one-dimensional vacancy chains as the charge cycle nears completion. [31] These findings underscore the versatility and significance of VEPALS in understanding complex defects and phenomena within battery materials.

These three complementary techniques, rarely explored within the battery field, offer a multifaceted perspective on the phase and defect evolution within LMO cathodes and their performance stability in aqueous electrolytes. Operating across a broad spectrum of length scales, they collectively reveal previously unseen features. Notably, VEPALS exposes the existence of cationic defects and sub-nanometer pores, providing insights at the atomic scale of the dynamic changes with cycling. TERS captures the presence of $Mn_3O_4$ at the surface and its subsequent dissolution during cycling with a resolution better than 20 nm. Simultaneously, TERS also highlights the accumulation and potential incorporation of $SO_4^{2-}$ ions, primarily at grain boundaries, while macro-Raman Spectroscopy focuses on the material's long-range order, depicting small changes at the micrometer scale. Our work contributes to a more profound understanding of the stability of LMO in aqueous electrolytes with the hope that this knowledge can be extrapolated to implement solutions toward longer stability in conventional organic electrolytes.

**Results & discussion**

Characterization of the as-deposited LMO films

LMO thin films were fabricated by Large Area Pulsed Laser Deposition with our well-established method of compensating Li losses with a multilayer deposition of $LiMn_2O_4$ parent compound and Li$_2$O acting as Li reservoir.[17,32–36] When grown on top of Pt-coated Si-chips, this method produces relatively rough polycrystalline layers with high phase purity. For this experiment, a batch of 300 nm thick LMO films was produced (for details, please see the Experimental section). SEM shows a cross-section of the layers that are dense, and rough (as expected) with well-connected grains with sizes of 250-300 nm (Fig. 1a). In line with our previous results,[17,35] the XRD patterns of the layer evidence the polycrystalline nature of our films with small quantities of Mn$_3$O$_4$ impurities. Despite the difficulty of performing more sophisticated approaches (Rietveld refinement) in this data, one can see for instance, that the most intense line that can be unambiguously assigned to Mn$_3$O$_4$ (2θ = 28.91º, I ≈ 40%) is almost not visible in the diffractogram. In contrast, the (400) reflection of LMO (2θ = 43.87º, I ≈ 33%) is identified with large intensity.

We further characterized the films with VEPALS. The average positron lifetime versus positron implantation energy is quite homogeneous along the film thickness (Fig. 1c). The deconvolution of the spectra (see Ref. [35], experimental methods and Supporting Information, section I) showed four main types of defects: small vacancy-like defects and their clusters and two families of sub-nm pores with spherical sizes $d_3$ ≈ 0.47 nm and $d_4$ ≈ 0.75 nm (calculated based on [37]). Overall, both pore families and vacancy-related defects exhibit a homogeneous distribution across the thickness of the film. Comparing the relative intensities, one can see that the vacancy-related defects dominate the average lifetime of the positrons, indicating the larger presence of vacancies than pores. These kinds of vacancies have also been observed by PALS in LiCoO$_2$ cathodes and have been ascribed to lithium vacancies and clusters of lithium vacancies, respectively.[27,30] However, we would rather be cautious over the assignment of the type of vacancy, see Supporting Information, section I for this discussion. Overall, these PALS measurements demonstrate the presence of cationic defects and pores in our films, which potentially play a role in suppressing and/or alleviating commonly observed Jahn−Teller distortions in LMO.[35,38]

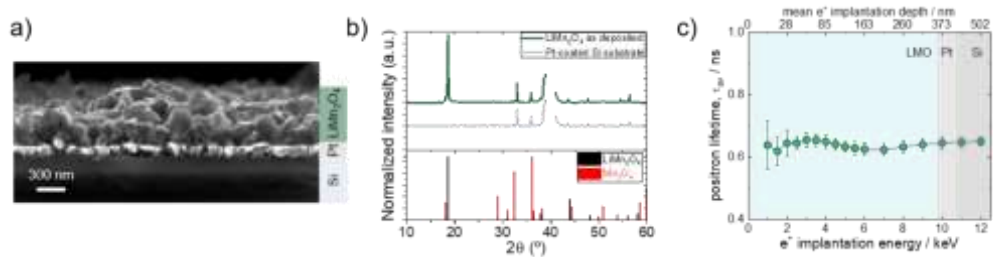

Figure 1 – Characterization of the as-deposited films. a) SEM cross-section of the as-deposited LMO films. b) XRD pattern of the LMO films and the bare Pt-coated Si substrate. c) Positron lifetime vs. implantation energy.

Regarding Raman characterization of the films, existing differences between the spectra obtained with macro-Raman Spectroscopy and TERS are considered relevant. On one hand, the low excitation energy, long acquisition times and very large laser spot size of the macro-Raman Spectroscopy assures that the signal arises from tens of thousands of grains, rather than from tens of grains interacting with a conventional Raman laser spot, given the grain sizes of our films, in the range of 250-300 nm. This allows detecting secondary phases and impurities extensively present in the films. The macro-Raman Spectroscopy of the as-deposited state films included in Fig. 2a shows the presence of mainly LiMn$_2$O$_4$ and impurities of Mn$_3$O$_4$,[39,40] in agreement with the XRD results. In addition, the presence of off-stoichiometric Li$_x$Mn$_2$O$_4$ phase cannot be discarded from the measured spectra,[39] suggesting the presence of vacancy defects in the bulk of the films. On the other hand, TERS measurements with higher surface sensitivity complement this information by assigning a large presence of $Mn_3O_4$ at the surface and sub-surface level. This is clearly shown in the average TERS spectra corresponding to a 500 x 500 $nm^2$ spectroscopy scan, Fig. 2b, where the main peak of $Mn_3O_4$ (located at ~660 $cm^{-1}$) is now more intense than the $LiMn_2O_4$ peaks, in stark contrast with the macro-Raman Spectroscopy measurements. The topography acquired during the TERS measurement, Fig. 2c, shows several grains, with very distinguishable grain boundaries. The TERS map, Fig. 2d, has been built in such a way that the green color scale represents the $LiMn_2O_4$ phase, linked to the intensity of the peak located at 580 $cm^{-1}$, and the blue color represents the $Mn_3O_4$, linked to the intensity of the peak at 660 $cm^{-1}$. While the $LiM_2O_4$ phase is more homogeneously distributed, the $Mn_3O_4$ phase seems to be located more at the bottom part of the image. The full power of the technique is revealed by superimposing the topography and the spectroscopy maps, Fig. 2e. Here, one can see that the $LiMn_2O_4$ is homogeneously distributed across the map, dominating the grains. In contrast, $Mn_3O_4$ is clearly present at the grain boundary level (all across the image) and extensively in some particular grains. For more details about how these maps are built and the demonstration of the TERS effect on our setup, the reader is referred to the Supporting Information, Section III.

Combining the different techniques, one can conclude that the LMO films are composed mainly of the electrochemically active spinel phase and impurities of $Mn_3O_4$, which are located primarily at the surface of the films and, especially, at the grain boundary level. The films also show the presence of vacancy-like defects and their clusters (probably at grain boundaries) across the film and to a lesser extent, two families of sub-nm pores.

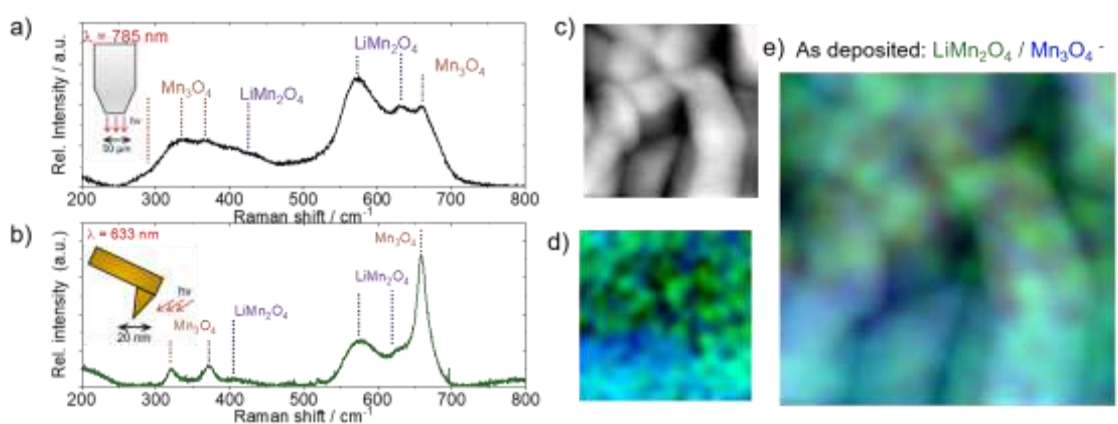

*Figure 2: Raman Spectroscopy and electrochemistry. a) Macro-Raman Spectroscopy. b) Average TERS spectrum of the TERS map in d). c) Topography obtained during the TERS mapping. Size 500 x 500 $nm^2$ d) TERS map, 500 x 500 $nm^2$. Green represents the $LiMn_2O_4$ phase and blue the $Mn_3O_4$ phase. e) Overlay of topography and TERS, 500 x 500 $nm^2$.*

<u>Characterization of the LMO films during cycling</u>

The fabricated polycrystalline LMO films were cycled as previously detailed elsewhere showing the same electrochemical and structural properties as those previously reported by our group.[17,35,41] Upon cycling in a sulphate-based aqueous electrolyte, the films show the expected two voltage plateaus and capacities of ~126 mAh·$g^{-1}$ after 300 cycles, with an estimated loss of 0.025% per cycle (Fig. 3a and inset). The electrochemical properties also evidence that the films are composed mainly of electrochemically active $LiMn_2O_4$ since its capacity approaches the practical capacity of $LiMn_2O_4$ that is commonly found in literature.[3]

We proceed here with a comprehensive exploration of the Raman signature evolution within our LMO films throughout the cycling process. Specifically, we analyzed films at 100% SoC (i.e., λ-$MnO_2$ phase) at different stages of cycling: After 1, 100 and 300 cycles, macro-Raman Spectroscopy allows us to scrutinize these changes. Fig. 3b illustrates a noteworthy trend: the primary peaks of the electrochemically active phases exhibit a relative intensity decrease concerning $Mn_3O_4$ inactive impurities. It is essential to note that our macroscopic Raman Spectroscopy approach, given its true average nature, is expected to detect all electrochemically active phases—namely, $LiMn_2O_4$, $Li_{0.5}Mn_2O_4$, and λ-$MnO_2$. This detection remains consistent with literature findings, as our capacities stay in the expected range of the practical capacity of the material.[3] This implies that a substantial portion of $Li^+$ ions resides within the lattice as $LiMn_2O_4$, even when the material is nominally at 100% State of Charge (SoC), as detected by our macro-Raman Spectroscopy measurements and other studies in the field.[42] The observed smearing of the Raman peaks associated with the electrochemically active phases can imply a gradual increase in film disorder and/or decrease of the crystalline quality. This phenomenon correlates with the rise in vacancy-like defects and their clusters, as identified through PALS in the next section. The enhanced disorder and reduced crystalline quality within electrochemically active phases during cycling may ultimately govern the long-term stability of these films in aqueous electrolytes, leading to a minimal capacity loss of only 0.025% per cycle.

In contrast, the $Mn_3O_4$ main mode appears to remain largely unaffected at the macroscopic level, emphasizing its inactive electrochemical role. However, a closer examination of the surface using Tip-Enhanced Raman Spectroscopy (TERS) mapping during cycling unveils two critical observations. Fig. 3c compares the average TERS spectra extracted from the corresponding maps. Firstly, there is a significant decrease in the $Mn_3O_4$ signal at the surface with cycling. This intriguing observation hints at the partial dissolution of these impurities, considering the presence of $Mn^{2+}$ within their structure.[3,43,44] Please note that we do not observe any obvious phase appearance with the dissolution of $Mn_3O_4$ seen at the surface. Secondly, the sulfate ions signal exhibits increased intensity and a blue shift, evolving from ~980 $cm^{-1}$ after 1 cycle to ~1007 $cm^{-1}$ after 300 cycles. This shift suggests a transition from absorbed $SO_4^{2-}$ ions to a scenario in which these adsorbates form stronger bonds with the film.

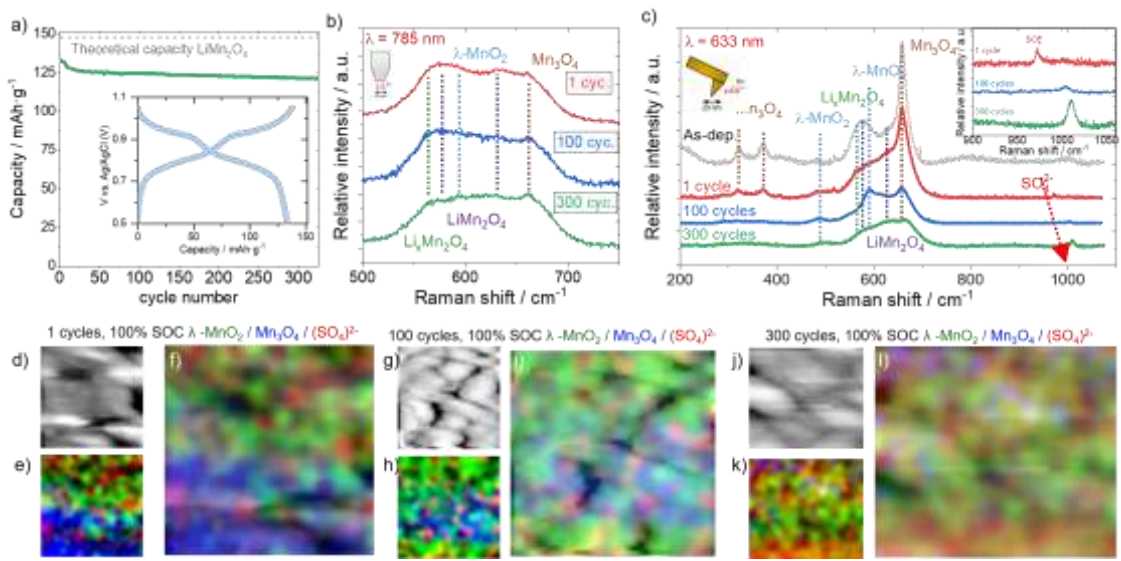

*Figure 3: Raman Spectroscopy of the cycled films. a) Electrochemical cycling of the LMO films. The inset shows an example of charge-discharge profile (cycle #50) b) Macro-Raman Spectroscopy. c) Average TERS spectrum of the*

TERS maps in e, h and k. For the film cycled 1 time, size 500 x 500 nm²: d) Topography, e) TERS map and f) overlay of both. For the film cycled 100 times, size 500 x 500 nm²: g) Topography, h) TERS map and i) overlay of both. For the film cycled 300 times, size 500 x 440 nm²: j) Topography, k) TERS map and l) overlay of both. All the measurements were done at full charge (SoC 100%).

Comparing the spatial distribution of $Mn_3O_4$ (blue) and sulfate signals (red) in films cycled 1x (Fig. 3d-f), 100x (Fig. 3g-i) and 300x (Fig. 3j-l) reveals a notable trend. In both cases, the proportion of maps dominated by $Mn_3O_4$ (blue) decreases with cycling (see Fig. 2e for as-deposited, Fig. 3f for 1, Fig. 3i for 100, and Fig. 3l for 300 cycles). By 300 cycles, the few visible blue spots are confined to certain grain boundaries. Moreover, both the 100x and 300x cycled maps exhibit the sulfate signal (red) predominantly at grain boundaries. To facilitate the identification, the reader is referred to the Supporting Information, Section III.

To complement our findings, VEPALS measurements were carried out to gain insights into defect evolution during cycling. Detailed descriptions of the deconvolution of the PALS signal into various components can be found in the Supporting Information, Section I. A critical observation emerges from the data analysis: our films' average positron lifetime ($\tau_{average}$) decreases from the first cycle onward, stabilizing after 100 and 300 cycles.

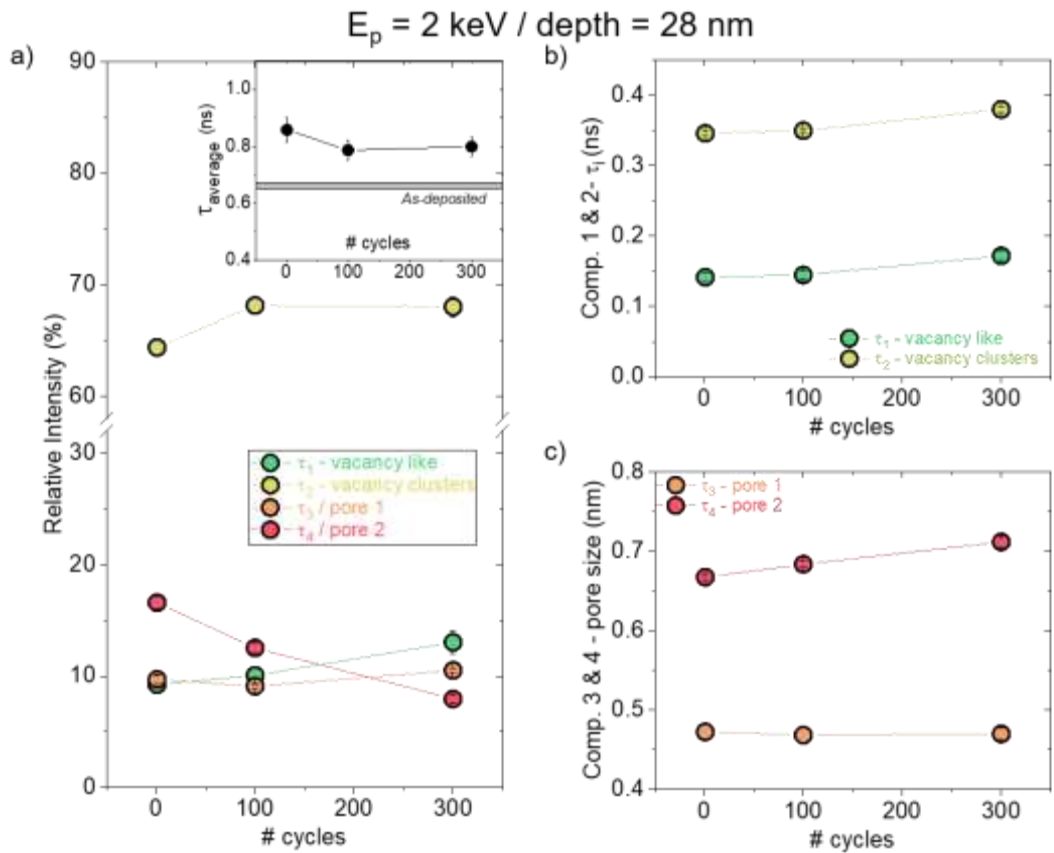

Figure 4: Positron annihilation lifetime spectroscopy of cycled films. a) Relative intensity of each type of defect as a function of cycling. Inset shows the positron average lifetime. b) Lifetime for components $\tau_1$ (vacancy-like defects) and $\tau_2$ (clusters of vacancies) as a function of size. d) Pore size derived from $\tau_3$ and $\tau_4$, resulting in $d_3$ and $d_4$, respectively, as a function of cycling.

This decline in $\tau_{average}$, directly linked to the free volume, can be attributed to the dynamic evolution of the electrolyte solution within grain boundaries and through pores within the film. Deconvolution of the different defect signals reveals that the most significant changes occur in $\tau_4$, representing a family of larger sub-nm pores. Its relative intensity, proportional to pore density, decreases from approximately 17% after 1 cycle to around 8% after 300 cycles. This decline could imply that a part of these characteristic pores might being filled, likely with sulfate ions, as suggested by our TERS measurements, or the pore size increase by coalescence to some extent. This shift in pore density redistributes the contribution of the other three components within the defects, which tend to become more significant in the signal. Specifically, the smaller vacancies ($\tau_1$) and their clusters ($\tau_2$) demonstrate an increase with cycling (Fig. 4b). This trend is mirrored in the large family of pores, $\tau_4$, which increases in size at a higher rate than vacancy clusters. In contrast, the smaller family of pores, $\tau_3$, appears unaffected by cycling. These changes, primarily the increases in $\tau_1$, $\tau_2$, and $\tau_4$ ($d_4$), suggest the dissolution of $Mn^{2+}$ (natively present in the $Mn_3O_4$ structure) into the solution, leaving behind larger defects.[44] Nevertheless, dissolution rate in pores is expected to be higher than due to vacancies and their clusters, given the higher free volume, as generally probed by PALS. It also aligns with the observed increase in film disorder, consistent with the smearing out of the Raman bands noted by macro-Raman Spectroscopy and TERS. For completeness, it is noteworthy that the substantial reduction in pore density ($\tau_4$) is even more pronounced at the film's bottom portion, essentially vanishing progressively with cycling (see Supporting Information, section II).

Our study aligns with recent research that underscores the dominant role of fast grain boundary and surface diffusion, surpassing grain diffusion by a staggering four orders of magnitude.[45] Mürter et al. have convincingly argued that the wide range of reported diffusion coefficients in the literature likely stems from underestimating the contribution of grain boundaries. In our work, we provide experimental evidence of the complex chemistry of grain boundaries during cycling, a critical factor that demands consideration. Notably, Mürter et al. identified residual electrolyte peaks in their X-ray diffraction (XRD) pattern. We propose an intriguing possibility: these residuals may, in fact, become integral parts of the material's structure after cycling, predominantly within the grain boundaries. This notion suggests that grain boundaries can actively evolve and adapt, potentially absorbing ions from the electrolyte as part of their transformation. These insights emphasize the need to account for the active role of grain boundaries when designing protective coatings, especially in high-voltage cathodes.

On the other hand, importantly, grain boundaries within $LiMn_2O_4$ are known to carry a positive charge and establish a space charge layer around them.[45,46] This inherent property makes them susceptible to ion adsorption from the electrolyte during cycling, warranting careful consideration in protective coating design. Understanding the behavior of grain boundaries in high-voltage cathodes represents a pivotal step towards the development of more sustainable and stable electrode materials.

The surface dilution of $Mn_3O_4$ during cycling, primarily restricted to a few grain boundaries after 300 cycles, stands in stark contrast to its behavior in organic electrolytes. In organic electrolytes, $Mn_3O_4$ tends to form at the $LiMn_2O_4$ grain surface upon charging to 4.3V vs. $Li/Li^+$, leading to partial irreversibility and cracking.[8,13] A recent work by Zhao et al. has shown that $Mn_3O_4$ nanowalls can be in-situ electrochemically oxidized in $Li_2SO_4$ electrolyte to form $Li_4Mn_5O_{12}$, a transformation with potential applications in supercapacitors.[47] Our TERS and macro-Raman Spectroscopy measurements do not conclusively reveal the formation of $Li_4Mn_5O_{12}$ or any other Mn-based polymorph upon cycling that could potentially correlate with the observed $Mn_3O_4$ dissolution. However, the stability in charge-discharge profiles suggests that this dissolution at the surface does not compromise the electrochemically active LMO electrode performance. It may even serve to protect the electrode from potential $Mn^{2+}$ dissolution.

Our findings regarding the presence of ionic species from the electrolyte anchored at the grain boundaries and surface not only contribute to unraveling the complexities of grain boundary chemistry but also align with recent efforts to incorporate lithium sulfate protective coatings in high-voltage cathodes, as documented in the literature.[48,49] These protective coatings can serve as a proactive shield against the evolving nature of grain boundaries, potentially mitigating the adverse effects of electrolyte interactions. Understanding the intricate interplay between grain boundaries and electrolytes is a pivotal step toward designing cathode materials that can perform well in the long term. Our work also holds broader implications for the stability of cathodes in aqueous electrolytes. For that, we argue that the presence of sub-nm pores and cationic vacancies can potentially help to mitigate the volume expansions in all-solid electrodes. These insights are not only pertinent to aqueous systems but also offer valuable lessons for enhancing the stability of cathodes in organic electrolytes and solid-state composites, where the behavior of grain boundaries and their interactions remain critical yet less explored.

**Conclusions**

Our research provides valuable structural and chemical insights across a spectrum of length scales for the defect evolution with cycling in LMO cathodes. Beginning at the sub-nanometer scale, we have identified vacancy-like defects and their clusters, with the latter emerging as the most prevalent defect within our films. These defects appear to grow in size with cycling, possibly due to the dissolution of $Mn^{2+}$ ions, introducing a degree of disorder into the films. Additionally, we have discerned the presence of two distinct families of sub-nanometer pores, notably, the larger pores ($\tau_4$) appear to reach a saturation point with cycling, hinting either at complete fill up of part of the pores in a gradual manner with electrolyte species or pore coalescence with cycling.

Zooming out to the nanoscale, our TERS measurements unveil that, unlike in organic electrolytes, $Mn_3O_4$ does not form at the surface during cycling in aqueous electrolytes. Instead, it undergoes dissolution and likely electrochemical transformation into another Mn-based phase, which remarkably maintains overall performance and stability. TERS also highlights the preferential adsorption of $SO_4^{2-}$ species at grain boundaries, with their bond strength appearing to intensify during cycling. In addition to the diminishing $Mn_3O_4$ signal, TERS and macro-Raman Spectroscopy suggest an increase in disorder within the electrochemically active Raman modes, aligning seamlessly with the observed enlargement of defects at the sub-nanometer scale. Notably, this marks the first successful utilization of TERS to investigate grain boundary effects in battery materials, revealing the strong bonding of electrolyte-derived species to the grain boundary core and its surroundings.

Expanding our observations to the macroscopic scale, macro-Raman Spectroscopy provides robust confirmation of the increased disorder within electrochemically active phases. Importantly, macro-Raman Spectroscopy maintains the overall intensity of the $Mn_3O_4$ impurities peaks. This finding validates our earlier TERS discovery that $Mn_3O_4$ impurities primarily dissolve at the surface. Notably, weak points, particularly grain boundaries, appear to be safeguarded by the absorption of sulfate ions.

Our study advances our understanding of this cathode behavior in aqueous electrolytes, holding great promise for the development of more sustainable and stable electrode materials. We hope this knowledge does not only pertain to aqueous systems but also extends towards enhancing the stability of cathodes in organic electrolytes. The dynamic nature of grain boundaries and their interactions with electrolytes must be taken into account in future protective coating designs.

**EXPERIMENTAL SECTION**

**Thin-Film Fabrication**. Commercially available ceramic targets of $LiMn_2O_4$ and $Li_2O$ (CODEX) were employed for the Larga-Area Pulsed Laser Deposition (LA PLD-5000, PVD Products), equipped with a Coherent (Lambda Physik) COMPex Pro 205 KrF excimer laser ($\lambda$ = 248 nm). Both targets were ablated sequentially following a multi-layer approach, as previously described.[35] The multi-layer sequence involved a consistent ratio of ablation pulses, specifically 4:3 between LMO (800 pulses per cycle) and $Li_2O$ (600 pulses per cycle). These depositions were conducted at 650°C and 20 mTorr $p_{O2}$, at a target-substrate distance of 90 mm, constant rotation and laser fluence of 1.3 J·cm$^{-2}$ inside the chamber. The substrates were Pt (80 nm)/ Ti (10 nm)/ $Si_3N_4$ (300 nm)/ $SiO_2$ (100 nm)/ Si (0.5 mm) chips of 1 x 1 cm$^2$, fabricated at the Institute of Microelectronics of Barcelona (IMB-CNM-CSIC).

**Structural characterization**. Scanning electron microscopy (SEM) was performed at IREC with a Zeiss Auriga (30 kV Gemini FESEM column and an in-lens detector). X-ray diffraction was performed thanks to an XRD Bruker-D8 Advance diffractometer in θ−2θ configuration between 10 and 60° (step size of 0.01°). The region around the main reflection of Pt was avoided during the measurement to protect the detector.

**Raman characterization**. Tip-Enhanced Raman Spectroscopy (TERS) measurements were performed on an Xplora Nano (HORIBA), with a 632 nm laser with p-polarization (parallel to the tip axis) and Au-coated OMNI TERS tips and 100x objective and a laser power of ~3 mW. Maps were acquired with a resolution of 10 nm per pixel to build 500 × 500 nm maps, plotted using a 2 px weight average. Acquisition times of 5 s per spectra and a 1200 g·mm$^{-1}$ grating were used. The protocol to ensure TERS signal consists of checking the TERS signal on a carbon nanotubes sample test before and after the experiment. In addition, TERS is demonstrated in our films, the reader is referred to the Supporting Information, section III, for more details about our TERS configuration. For the macro-Raman Spectroscopy measurements, Raman scattering spectra were measured using iHR320 Horiba Jobin-Yvon monochromator coupled with i-DUS CCD from Andor. NIR solid-state laser (785 nm, ~80 W/cm$^2$) was used as excitation source. The measurements were performed in backscattering configuration through the specific probes developed in IREC. The spectral position was corrected using monocrystalline Si as a reference by imposing the main Raman peak to 520 cm$^{-1}$.

**Variable energy positron annihilation lifetime spectroscopy**. VEPALS was carried out at the Mono-energetic Positron Source (MePS) beamline at HZDR (Germany) [50] using a digital lifetime $CrBr_3$ scintillator detector. This was coupled to a Hamamatsu R13089-100 PMT μ-metal shielded and housed inside an Au solid casing. We used homemade software employing a SPDevices ADQ14DC-2X digitizer with 14-bit vertical and 2GS/s horizontal resolutions [51]. The time resolution function was down to 0.250 ns. The spectra analysis consisted of a resolution function of two Gaussian with distinct intensities and shifts depending on the positron implantation energy, $E_p$. All spectra contained at least $1\cdot10^7$ counts. Typical lifetime spectrum N(t) is described by $N(t)=\Sigma_i (1/\tau_i) I_i \exp(-t/\tau_i)$, where $\tau_i$ and $I_i$ are the positron lifetime and intensity of the i-th component, respectively ($\Sigma I_i=1$). All the spectra were deconvoluted using a non-linear least-squares fitting method employed within the fitting software package PALSfit.[52] It consisted of 5 discrete lifetime components, which directly show localized annihilation at 2 different defect types (sizes; $\tau_1$ and $\tau_2$), i.e., small vacancy-like defect and their agglomerations or clusters. 3rd and 4th lifetime components ($\tau_3$ and $\tau_4$) are two pore populations of diameters $d_3$ and $d_4$. The 5th component (not shown) is residual and originates from ortho-positronium annihilation in vacuum and pore networks. The pore size was derived using the Wada and Hyodo shape-free model.[37] The positron lifetime and its intensity have been probed as a function of positron implantation energy $E_p$ or equivalently, implantation depth or film thickness. Positrons have been accelerated and monoenergetically implanted into samples in the range between $E_p$ = 1-12 keV for depth profiling. A mean positron implantation depth was approximated using a simple material density ($\rho$ = 4.02 g·cm$^{-3}$) dependent formula: $<z>=36/\rho\cdot E_p^{1.62}$.[53] The average positron lifetime $\tau_{average}$ is defined as $\tau_{average} = \sum_i \tau_i \cdot I_i$.

**Electrochemical measurements.** For electrochemical measurements, we established electrical contacts with the LMO films by accessing the exposed Pt regions at the sample edges, securely embedding them within a durable dual-component resin. This resin was applied thinly enough to facilitate mechanically stable TERS and PALS characterizations. The electrochemical measurements were conducted employing an open cell assembly, wherein a substantial portion of the film's surface remained exposed, facilitated by the resin's coverage. The exposed surface area typically ranged from 30 to 40 mm$^2$. This configuration allowed for the rapid quenching of the electrochemical cell, subsequent immersion of the film in DI water, and fast drying using a nitrogen gun for storage within a protective atmosphere. The electrochemical experiments encompassed a voltage window spanning from 0.5 to 1.05 V versus Ag/AgCl (3M KCl), employing a Pt mesh as the counter electrode.

**Acknowledgments**

This project has received funding from the European Union's Horizon 2020 research and innovation program under grant agreement No 824072 (HARVESTORE), from the European Regional Development Fund under the FEDER Catalonia Operative Programme 2014-2020 (FEM-IoT, 001-P-00166) and the "Generalitat de Catalunya" (2021 SGR 00750, NANOEN, and 2021 SGR 01286). J. C. G.- R. acknowledges the financial support provided by the European Union's Horizon 2020 research and innovation program under the Marie Skłodowska-Curie Grant Agreement No. 801342 (Tecniospring INDUSTRY), as well as by the Agency for Business Competitiveness of the Government of Catalonia. Parts of this research were carried out at ELBE at the Helmholtz-Zentrum Dresden - Rossendorf e. V., a member of the Helmholtz Association. We would like to thank the facility staff for their assistance.

**REFERENCES**

[1] J.C. Hunter, Preparation of a new crystal form of manganese dioxide: λ-MnO2, J. Solid State Chem. 39 (1981) 142–147. https://doi.org/10.1016/0022-4596(81)90323-6.

[2] M.M. Thackeray, W.I.F. David, P.G. Bruce, J.B. Goodenough, Lithium Insertion Into Manganese Spinels, Mater. Res. 18 (1983) 461–472.

[3] Y. Huang, Y. Dong, S. Li, J. Lee, C. Wang, Z. Zhu, W. Xue, Y. Li, J. Li, Lithium Manganese Spinel Cathodes for Lithium-Ion Batteries, Adv. Energy Mater. 11 (2021) 1–21. https://doi.org/10.1002/aenm.202000997.

[4] W. Zhang, D. Seo, T. Chen, L. Wu, M. Topsakal, Y. Zhu, D. Lu, G. Ceder, F. Wang, Kinetic pathways of ionic transport in fast-charging lithium titanate, Science (80-. ). 367 (2020) 1030–1034. https://doi.org/10.1126/science.aax3520.

[5] X. Hou, X. Liu, H. Wang, X. Zhang, J. Zhou, M. Wang, Specific countermeasures to intrinsic capacity decline issues and future direction of LiMn2O4 cathode, Energy Storage Mater. 57 (2023) 577–606. https://doi.org/10.1016/j.ensm.2023.02.015.

[6] D.H. Jang, Y.J. Shin, S.M. Oh, Dissolution of Spinel Oxides and Capacity Losses in 4 V Li / Li x Mn2 O 4 Cells, J. Electrochem. Soc. 143 (1996) 2204–2211. https://doi.org/10.1149/1.1836981.

[7] L.-F. Wang, C.-C. Ou, K.A. Striebel, J.-S. Chen, Study of Mn Dissolution from LiMn[sub 2]O[sub 4] Spinel Electrodes Using Rotating Ring-Disk Collection Experiments, J. Electrochem. Soc. 150 (2003) A905. https://doi.org/10.1149/1.1577543.

[8] D. Tang, Y. Sun, Z. Yang, L. Ben, L. Gu, X. Huang, Surface structure evolution of LiMn2O4 cathode material upon charge/discharge, Chem. Mater. 26 (2014) 3535–3543. https://doi.org/10.1021/cm501125e.

[9] X. Gao, Y.H. Ikuhara, C.A.J. Fisher, R. Huang, A. Kuwabara, H. Moriwake, K. Kohama, Y. Ikuhara, Oxygen loss and surface degradation during electrochemical cycling of lithium-ion battery cathode material LiMn2O4, J. Mater. Chem. A. 7 (2019) 8845–8854. https://doi.org/10.1039/c8ta08083f.

[10] M. Hirayama, H. Ido, K. Kim, W. Cho, Dynamic Structural Changes at LiMn 2 O 4 / Electrolyte Interface, J. Am. Chem. Soc. 132 (2010) 15268–15276.

[11] R. Huang, Y.H. Ikuhara, T. Mizoguchi, S.D. Findlay, A. Kuwabara, C.A.J. Fisher, H. Moriwake, H. Oki, T. Hirayama, Y. Ikuhara, Oxygen-vacancy ordering at surfaces of lithium manganese(III,IV) oxide spinel nanoparticles, Angew. Chemie - Int. Ed. 50 (2011) 3053–3057. https://doi.org/10.1002/anie.201004638.

[12] D. Aurbach, M.D. Levi, K. Gamulski, B. Markovsky, G. Salitra, E. Levi, U. Heider, L. Heider, R. Oesten, Capacity fading of LixMn2O4 spinel electrodes studied by XRD and electroanalytical techniques, J. Power Sources. 81–82 (1999) 472–479. https://doi.org/10.1016/S0378-7753(99)00204-9.

[13] T. Liu, A. Dai, J. Lu, Y. Yuan, Y. Xiao, L. Yu, M. Li, J. Gim, L. Ma, J. Liu, C. Zhan, L. Li, J. Zheng, Y. Ren, T. Wu, R. Shahbazian-Yassar, J. Wen, F. Pan, K. Amine, Correlation between manganese dissolution and dynamic phase stability in spinel-based lithium-ion battery, Nat. Commun. 10 (2019) 4721. https://doi.org/10.1038/s41467-019-12626-3.

[14] Q. Qu, L. Fu, X. Zhan, D. Samuelis, J. Maier, L. Li, S. Tian, Z. Li, Y. Wu, Porous LiMn 2O 4 as cathode material with high power and excellent cycling for aqueous rechargeable lithium batteries, Energy Environ. Sci. 4 (2011) 3985–3990. https://doi.org/10.1039/c0ee00673d.

[15] L. Tian, A. Yuan, Electrochemical performance of nanostructured spinel LiMn2O4 in different aqueous electrolytes, J. Power Sources. 192 (2009) 693–697. https://doi.org/10.1016/j.jpowsour.2009.03.002.

[16] O. Hanna, D. Malka, S. Luski, T. Brousse, D. Aurbach, Aqueous Energy Storage Device Based on LiMn 2 O 4 ( Spinel ) Positive Electrode and Anthraquinone-Modi fi ed Carbon-Negative Electrode, 4 (2019) 1–9. https://doi.org/10.1002/ente.201900589.

[17] M. Fehse, R. Trócoli, E. Ventosa, E. Hernández, A. Sepúlveda, A. Morata, A. Tarancón, Ultrafast Dischargeable LiMn 2 O 4 Thin-Film Electrodes with Pseudocapacitive Properties for Microbatteries, ACS Appl. Mater. Interfaces. 9 (2017) 5295–5301. https://doi.org/10.1021/acsami.6b15258.

[18] W. Tang, S. Tian, L.L. Liu, L. Li, H.P. Zhang, Y.B. Yue, Y. Bai, Y.P. Wu, K. Zhu, Nanochain LiMn2O4 as ultra-fast cathode material for aqueous rechargeable lithium batteries, Electrochem. Commun. 13 (2011) 205–208. https://doi.org/10.1016/j.elecom.2010.12.015.

[19] M. Huang, X. Wang, X. Liu, L. Mai, Fast Ionic Storage in Aqueous Rechargeable Batteries: From Fundamentals to Applications, Adv. Mater. 34 (2022) 2105611. https://doi.org/10.1002/ADMA.202105611.

[20] T. Schmid, L. Opilik, C. Blum, R. Zenobi, Nanoscale chemical imaging using tip-enhanced raman spectroscopy: A critical review, Angew. Chemie - Int. Ed. 52 (2013) 5940–5954. https://doi.org/10.1002/anie.201203849.

[21] P. Verma, Tip-Enhanced Raman Spectroscopy: Technique and Recent Advances, Chem. Rev. 117 (2017) 6447–6466. https://doi.org/10.1021/acs.chemrev.6b00821.

[22] J. Langer, D.J. de Aberasturi, J. Aizpurua, R.A. Alvarez-Puebla, B. Auguié, J.J. Baumberg, G.C. Bazan, S.E.J. Bell, A. Boisen, A.G. Brolo, J. Choo, D. Cialla-May, V. Deckert, L. Fabris, K. Faulds, F. Javier García de Abajo, R. Goodacre, D. Graham, A.J. Haes, C.L. Haynes, C. Huck, T. Itoh, M. Käll, J. Kneipp, N.A. Kotov, H. Kuang, E.C. Le Ru, H.K. Lee, J.F. Li, X.Y. Ling, S.A. Maier, T. Mayerhöfer, M. Moskovits, K. Murakoshi, J.M. Nam, S. Nie, Y. Ozaki, I. Pastoriza-Santos, J. Perez-Juste, J. Popp, A. Pucci, S. Reich, B. Ren, G.C. Schatz, T. Shegai, S. Schlücker, L.L. Tay, K. George Thomas, Z.Q. Tian, R.P. van Duyne, T. Vo-Dinh, Y. Wang, K.A. Willets, C. Xu, H. Xu, Y. Xu, Y.S. Yamamoto, B. Zhao, L.M. Liz-Marzán, Present and future of surface-enhanced Raman scattering, ACS Nano. 14 (2020) 28–117. https://doi.org/10.1021/acsnano.9b04224.

[23] Y. Cao, M. Sun, Tip-enhanced Raman spectroscopy, Rev. Phys. 8 (2022) 100067. https://doi.org/10.1016/j.revip.2022.100067.

[24] J. Nanda, G. Yang, T. Hou, D.N. Voylov, X. Li, R.E. Ruther, M. Naguib, K. Persson, G.M. Veith, A.P. Sokolov, Unraveling the Nanoscale Heterogeneity of Solid Electrolyte Interphase Using Tip-Enhanced Raman Spectroscopy, Joule. 3 (2019) 2001–2019. https://doi.org/10.1016/j.joule.2019.05.026.

[25] G. Yang, X. Li, Y. Cheng, M. Wang, D. Ma, A. Sokolov, S. Kalinin, G. Veith, J. Nanda, Distilling Nanoscale Heterogeneity of Amorphous Silicon using Tip-

enhanced Raman Spectroscopy (TERS) via Multiresolution Manifold Learning, Nat. Commun. (2020). https://doi.org/10.21203/rs.3.rs-38466/v1.

[26] T.X. Huang, S.C. Huang, M.H. Li, Z.C. Zeng, X. Wang, B. Ren, Tip-enhanced Raman spectroscopy: Tip-related issues, Anal. Bioanal. Chem. 407 (2015) 8177–8195. https://doi.org/10.1007/s00216-015-8968-8.

[27] J. Nokelainen, B. Barbiellini, J. Kuriplach, S. Eijt, R. Ferragut, X. Li, V. Kothalawala, K. Suzuki, H. Sakurai, H. Hafiz, K. Pussi, F. Keshavarz, A. Bansil, Identifying Redox Orbitals and Defects in Lithium-Ion Cathodes with Compton Scattering and Positron Annihilation Spectroscopies: A Review, Condens. Matter. 7 (2022) 1–18. https://doi.org/10.3390/condmat7030047.

[28] P.G. Coleman, S.C. Sharma, L.M. Diana, Positron annihilation : proceedings of the sixth International Conference on Positron Annihilation, the University of Texas at Arlington, April 3-7, 1982, North-Holland Pub. Co., 1982. https://inis.iaea.org/search/search.aspx?orig_q=RN:14763881 (accessed October 6, 2023).

[29] M. Charlton, J.W. Humberston, Positron Physics, Cambridge University Press, 2000. https://doi.org/10.1017/CBO9780511535208.

[30] G. Pagot, V. Toso, B. Barbiellini, R. Ferragut, V. Di Noto, Positron annihilation spectroscopy as a diagnostic tool for the study of licoo2 cathode of lithium-ion batteries, Condens. Matter. 6 (2021) 1–11. https://doi.org/10.3390/condmat6030028.

[31] G. Klinser, H. Kren, S. Koller, R. Würschum, Operando monitoring of charging-induced defect formation in battery electrodes by positrons, Appl. Phys. Lett. 114 (2019). https://doi.org/10.1063/1.5081668.

[32] M. Fehse, R. Trócoli, E. Hernández, E. Ventosa, A. Sepúlveda, A. Morata, A. Tarancón, An innovative multi-layer pulsed laser deposition approach for LiMn2O4 thin film cathodes, Thin Solid Films. 648 (2018) 108–112. https://doi.org/10.1016/j.tsf.2018.01.015.

[33] R. Trócoli, A. Morata, C. Erinmwingbovo, F. La Mantia, A. Tarancón, Self-discharge in Li-ion aqueous batteries: A case study on LiMn2O4, Electrochim. Acta. 373 (2021) 137847. https://doi.org/10.1016/j.electacta.2021.137847.

[34] V. Siller, J.C. Gonzalez-Rosillo, M. Nuñez Eroles, M. Stchakovsky, R. Arenal, A. Morata, A. Tarancón, Safe extended-range cycling of Li4Ti5O12-based anodes for ultra-high capacity thin film batteries, Mater. Today Energy. 25 (2022) 100979. https://doi.org/10.1016/j.mtener.2022.100979.

[35] V. Siller, J.C. Gonzalez-Rosillo, M. Nuñez Eroles, F. Baiutti, M.O. Liedke, M. Butterling, A.G. Attallah, A. Wagner, A. Morata, A. Tarancon, Nanoscaled LiMn2O4 for Extended Cycling Stability in the 3 V Plateau, ACS Appl. Mater. Interfaces. 14 (2022) 33438–33446. https://doi.org/https://doi.org/10.1021/acsami.2c10798.

[36] C. Erinmwingbovo, V. Siller, M. Nuñez, R. Trócoli, D. Brogioli, A. Tarancón, A. Morata, F. La Mantia, Effect of Film Thickness on the Kinetics of Lithium Insertion in LiMn2O4 Films Made by Multilayer Pulsed Laser Deposition for Thin-Film All-Solid-State Battery Cathode Materials**, ChemElectroChem. 202200759 (2023) 1–7. https://doi.org/10.1002/celc.202200759.

[37] K. Wada, T. Hyodo, A simple shape-free model for pore-size estimation with positron annihilation lifetime spectroscopy, J. Phys. Conf. Ser. 443 (2013).

https://doi.org/10.1088/1742-6596/443/1/012003.

[38] R. Arabolla Rodríguez, N. Della Santina Mohallem, M. Avila Santos, D.A. Sena Costa, L. Andrey Montoro, Y. Mosqueda Laffita, L.A. Tavera Carrasco, E.L. Perez-Cappe, Unveiling the role of Mn-interstitial defect and particle size on the Jahn-Teller distortion of the LiMn2O4 cathode material, J. Power Sources. 490 (2021). https://doi.org/10.1016/j.jpowsour.2021.229519.

[39] A. Paolone, A. Sacchetti, T. Corridoni, P. Postorino, R. Cantelli, G. Rousse, C. Masquelier, MicroRaman spectroscopy on LiMn2O4: Warnings on laser-induced thermal decomposition, Solid State Ionics. 170 (2004) 135–138. https://doi.org/10.1016/j.ssi.2004.02.002.

[40] Q. Shi, Y. Takahashi, J. Akimoto, I.C. Stefan, D.A. Scherson, In situ Raman scattering measurements of a LiMn2O4 single crystal microelectrode, Electrochem. Solid-State Lett. 8 (2005) 6–10. https://doi.org/10.1149/1.2030507.

[41] A. Morata, V. Siller, F. Chiabrera, M. Nuñez, R. Trocoli, M. Stchakovsky, A. Tarancón, Operando probing of Li-insertion into LiMn2O4 cathodes by spectroscopic ellipsometry, J. Mater. Chem. A. 8 (2020) 11538–11544. https://doi.org/10.1039/C9TA12723B.

[42] K. Dokko, Q. Shi, I.C. Stefan, D.A. Scherson, In Situ Raman Spectroscopy of Single Microparticle Li +-Intercalation Electrodes, J. Phys. Chem. B. 107 (2003) 12549–12554. https://doi.org/10.1021/jp034977c.

[43] S.A. Beknalkar, A.M. Teli, T.S. Bhat, K.K. Pawar, S.S. Patil, N.S. Harale, J.C. Shin, P.S. Patil, Mn3O4 based materials for electrochemical supercapacitors: Basic principles, charge storage mechanism, progress, and perspectives, J. Mater. Sci. Technol. 130 (2022) 227–248. https://doi.org/10.1016/j.jmst.2022.03.036.

[44] T.À.H. Wu, L.À.H. Yen, Y.À.Q. Lin, Defect regulated spinel Mn 3 O 4 obtained by glycerol – assisted method for high – energy – density aqueous zinc – ion batteries, J. Colloid Interface Sci. 625 (2022) 354–362. https://doi.org/https://doi.org/10.1016/j.jcis.2022.06.033.

[45] J. Mürter, S. Nowak, E. Hadjixenophontos, Y. Joshi, G. Schmitz, Grain boundary transport in sputter-deposited nanometric thin films of lithium manganese oxide, Nano Energy. 43 (2018) 340–350. https://doi.org/10.1016/j.nanoen.2017.11.038.

[46] C. Schwab, A. Höweling, A. Windmüller, J. Gonzalez-Julian, S. Möller, J.R. Binder, S. Uhlenbruck, O. Guillon, M. Martin, Bulk and grain boundary Li-diffusion in dense LiMn2O4 pellets by means of isotope exchange and ToF-SIMS analysis, Phys. Chem. Chem. Phys. 21 (2019) 26066–26076. https://doi.org/10.1039/c9cp05128g.

[47] N. Zhao, H. Fan, M. Zhang, J. Ma, C. Wang, A.K. Yadav, H. Li, X. Jiang, X. Cao, Beyond intercalation-based supercapacitors: The electrochemical oxidation from Mn3O4 to Li4Mn5O12 in Li2SO4 electrolyte, Nano Energy. 71 (2020) 104626. https://doi.org/10.1016/j.nanoen.2020.104626.

[48] H. Sclar, S. Maiti, R. Sharma, E.M. Erickson, J. Grinblat, R. Raman, M. Talianker, M. Noked, A. Kondrakov, B. Markovsky, D. Aurbach, Improved Electrochemical Behavior and Thermal Stability of Li and Mn-Rich Cathode Materials Modified by Lithium Sulfate Surface Treatment, Inorganics. 10 (2022). https://doi.org/10.3390/inorganics10030039.

[49] M. Heidbüchel, T. Schultz, T. Placke, M. Winter, N. Koch, R. Schmuch, A.

Gomez-Martin, Enabling Aqueous Processing of Ni-Rich Layered Oxide Cathode Materials by Addition of Lithium Sulfate, ChemSusChem. 16 (2023). https://doi.org/10.1002/cssc.202202161.

[50] A. Wagner, M. Butterling, M.O. Liedke, K. Potzger, R. Krause-Rehberg, Positron annihilation lifetime and Doppler broadening spectroscopy at the ELBE facility, in: AIP Conf. Proc., 2018: p. 040003. https://doi.org/10.1063/1.5040215.

[51] E. Hirschmann, M. Butterling, U. Hernandez Acosta, M.O. Liedke, A.G. Attallah, P. Petring, M. Görler, R. Krause-Rehberg, A. Wagner, A new system for real-time data acquisition and pulse parameterization for digital positron annihilation lifetime spectrometers with high repetition rates, J. Instrum. 16 (2021) P08001. https://doi.org/10.1088/1748-0221/16/08/P08001.

[52] J. V. Olsen, P. Kirkegaard, N.J. Pedersen, M. Eldrup, PALSfit: A new program for the evaluation of positron lifetime spectra, Phys. Status Solidi. 4 (2007) 4004–4006. https://doi.org/10.1002/pssc.200675868.

[53] J. Dryzek, P. Horodek, GEANT4 simulation of slow positron beam implantation profiles, Nucl. Instruments Methods Phys. Res. Sect. B Beam Interact. with Mater. Atoms. 266 (2008) 4000–4009. https://doi.org/10.1016/j.nimb.2008.06.033.